\documentclass{eptcs-modified}
 \providecommand{\event}{} 
 \usepackage{breakurl}             
\usepackage{amsmath}
\usepackage{amssymb}
\usepackage{amsthm}
\usepackage{amscd}
\usepackage{graphics}

\long\def\greybox#1{%
    \newbox\contentbox%
    \newbox\bkgdbox%
    \setbox\contentbox\hbox to \hsize{%
        \vtop{
            \kern\columnsep
            \hbox to \hsize{%
                \kern\columnsep%
                \advance\hsize by -2\columnsep%
                \setlength{\textwidth}{\hsize}%
                \vbox{
                    \parskip=\baselineskip
                    \parindent=0bp
                    #1
                }%
                \kern\columnsep%
            }%
            \kern\columnsep%
        }%
    }%
    \setbox\bkgdbox\vbox{
        \pdfliteral{0.85 0.85 0.85 rg}
        \hrule width  \wd\contentbox %
               height \ht\contentbox %
               depth  \dp\contentbox
        \pdfliteral{0 0 0 rg}
    }%
    \wd\bkgdbox=0bp%
    \vbox{\hbox to \hsize{\box\bkgdbox\box\contentbox}}%
    \vskip\baselineskip%
}

\title{The Descriptive Theory of Represented Spaces}
\author{
Arno Pauly
\institute{Clare College\\ University of Cambridge, United Kingdom}
\email{Arno.Pauly@cl.cam.ac.uk}
}
\def\titlerunning{Descriptive Theory of Represented Spaces}
\def\authorrunning{A. Pauly}

\begin{document}
\theoremstyle{definition}
\newtheorem{theorem}{Theorem}
\newtheorem{definition}[theorem]{Definition}
\newtheorem{problem}[theorem]{Problem}
\newtheorem{assumption}[theorem]{Assumption}
\newtheorem{corollary}[theorem]{Corollary}
\newtheorem{proposition}[theorem]{Proposition}
\newtheorem{lemma}[theorem]{Lemma}
\newtheorem{observation}[theorem]{Observation}
\newtheorem{fact}[theorem]{Fact}
\newtheorem{question}[theorem]{Open Question}
\newtheorem{conjecture}[theorem]{Conjecture}
\newtheorem{example}[theorem]{Example}
\newcommand{\dom}{\operatorname{dom}}
\newcommand{\id}{\textnormal{id}}
\newcommand{\Cantor}{{\{0, 1\}^\mathbb{N}}}
\newcommand{\Baire}{{\mathbb{N}^\mathbb{N}}}
\newcommand{\Lev}{\textnormal{Lev}}
\newcommand{\hide}[1]{}
\newcommand{\mto}{\rightrightarrows}
\newcommand{\uint}{{[0, 1]}}
\newcommand{\bft}{\mathrm{BFT}}
\newcommand{\lbft}{\textnormal{Linear-}\mathrm{BFT}}
\newcommand{\pbft}{\textnormal{Poly-}\mathrm{BFT}}
\newcommand{\sbft}{\textnormal{Smooth-}\mathrm{BFT}}
\newcommand{\ivt}{\mathrm{IVT}}
\newcommand{\cc}{\textrm{CC}}
\newcommand{\lpo}{\textrm{LPO}}
\newcommand{\llpo}{\textrm{LLPO}}
\newcommand{\aou}{AoU}
\newcommand{\Ctwo}{C_{\{0, 1\}}}
\newcommand{\name}[1]{\textsc{#1}}
\newcommand{\C}{\textrm{C}}
\newcommand{\ic}[1]{\textrm{C}_{\sharp #1}}
\newcommand{\xc}[1]{\textrm{XC}_{#1}}
\newcommand{\me}{\name{P}.~}
\newcommand{\etal}{et al.~}
\newcommand{\eval}{\operatorname{eval}}
\newcommand{\Sierp}{Sierpi\'nski }
\newcommand{\isempty}{\operatorname{IsEmpty}}
\newcommand{\spec}{\textrm{Spec}}

\newcounter{saveenumi}
\newcommand{\seti}{\setcounter{saveenumi}{\value{enumi}}}
\newcommand{\conti}{\setcounter{enumi}{\value{saveenumi}}}

\maketitle

\begin{abstract}
This is a survey on the ongoing development of a descriptive theory of represented spaces, which is intended as an extension of both classical and effective descriptive set theory to deal with both sets and functions between represented spaces. Most material is from work-in-progress, and thus there may be a stronger focus on projects involving the author than an objective survey would merit.
\end{abstract}

\section{Introduction}
On the surface, the emergent \emph{descriptive theory of represented spaces} is the extension of both classical \cite{kechris} and effective \cite{moschovakis} descriptive set theory from its usual setting of Polish spaces to the much larger class of represented spaces. There is far more to it, though: By extending the setting, suddenly connections to areas such as category theory (with a certain topos-theoretic flavour) and computability-theoretic degree theory appear at a fundamental level.

Initial results do inspire some hope that this new theory can help to solve existing open problems, elucidate further the core aspects of known theorems and provide both new questions and theorems in its own right. The present text is meant both as a survey of existing work relevant to (or part of) the descriptive theory of represented spaces, and to outline some exciting directions of further research.

\subsection{The first step: Quasi-Polish spaces}
A suitable formalism for extending core results from descriptive set theory to a larger class of spaces was provided by \name{de Brecht} in the quasi-Polish spaces \cite{debrecht6}. These are Smyth-complete quasi-metrizable separable spaces, and in particular contain all the $\omega$-continuous domains, which have seen some extensions of descriptive set theory before, in the work of \name{Selivanov} \cite{selivanov6}.

\name{de Brecht} \cite{debrecht6,debrecht5} and others \cite{schlicht} have demonstrated that quasi-Polish spaces indeed are a suitable setting for pursuing descriptive set theory, without many fundamental changes. One relevant change that does occur is the necessity to slightly alter the usual definition of the Borel hierarchy in order to keep its desired properties, which was found out by \name{Selivanov} \cite{selivanov3}.

Quasi-Polish spaces do have a nice characterization as represented spaces: They are precisely the countably based admissible spaces with a total Baire-space representation. This characterization may often be easier to work with than that via a Smyth-complete quasi-metric, and foreshadows the appearances of notions from the theory of represented spaces.

\subsection{Category theory enters the fray}
In its traditional construction, descriptive set theory refers to at least two very distinct types of objections: Polish spaces, and point classes. Apart from the lowest level, point classes are generally not metrizable. Thus, quite similar operations such as the maximum of two real numbers and the union of two $\Sigma^0_n$-sets cannot be seen in a common framework without extending the setting.

The category of represented spaces is not only cartesian closed, but has sufficient additional structure that given a represented space, each instance of the usual point classes is a represented space again. Note that this does require to work with the entire category of represented spaces, and to disavow the common restriction to admissibly represented spaces.

Once viewed in their natural place in the category of represented spaces, the usual constructions from descriptive set theory are closely tight to certain endofunctors. As observed by the author and \name{de Brecht} in \cite{pauly-descriptive}, it is even possible to develop parts of the theory without referring to any specific properties of the category of represented spaces, but to pursue \emph{synthetic descriptive set theory} as the investigation of certain constructions on endofunctors in suitable cartesian closed categories. This is very similar to and in fact an extension of the realization that a significant part of topology can be developed in a suitable cartesian closed category in the form of \emph{synthetic topology}, outlined by \name{Escard\'o} in \cite{escardo}.

The fundamental notions of synthetic descriptive set theory (as far as already identified) are described in Section \ref{section:synthetic}. A case study follows in Section \ref{sec:borel}.

\subsection{From spaces to points}
It has been observed that the computable counterparts to theorems in classical analysis holding almost everywhere actually make meaningful statements about points, namely about the random points. Instances of this are present in \cite{nies2,nies3}. A similar phenomenon happens for dimension, too: The \emph{point degree spectrum} of a space as introduced by \name{Kihara} and the author in \cite{pauly-kihara-arxiv} is the collection of Medvedev degrees of its points, and characterizes a space up to countably computable isomorphisms. This both means that a single point can carry complexity (before relativization), and provides a convenient setting to explore properties of spaces linked to properties of names -- potentially including randomness in settings beyond computable metric spaces. Section \ref{sec:pointspectra} introduces the fundamental definitions and results obtained so far.

\section{A short introduction to represented spaces}
We briefly present some fundamental concepts on represented spaces following \cite{pauly-synthetic-arxiv}, to which the reader shall also be referred for a more detailed presentation. The concept behind represented spaces essentially goes back to \name{Weihrauch} and \name{Kreitz} \cite{kreitz}, the name may have first been used by \name{Brattka} \cite{brattka13}. A \emph{represented space} is a pair $\mathbf{X} = (X, \delta_X)$ of a set $X$ and a partial surjection $\delta_X : \subseteq \Baire \to X$. A function between represented spaces is a function between the underlying sets. For $f : \mathbf{X} \to \mathbf{Y}$ and $F : \subseteq \Baire \to \Baire$, we call $F$ a realizer of $f$ (notation $F \vdash f$), iff $\delta_Y(F(p)) = f(\delta_X(p))$ for all $p \in \dom(f\delta_X)$, i.e.~if the following diagram commutes:
 $$\begin{CD}
\Baire @>F>> \Baire\\
@VV\delta_\mathbf{X}V @VV\delta_\mathbf{Y}V\\
\mathbf{X} @>f>> \mathbf{Y}
\end{CD}$$
A map between represented spaces is called computable (continuous), iff it has a computable (continuous) realizer. Similarly, we call a point $x \in \mathbf{X}$ computable, iff there is some computable $p \in \Baire$ with $\delta_\mathbf{X}(p) = x$. A priori, the notion of a continuous map between represented spaces and a continuous map between topological spaces are distinct and should not be confused!

We arrive at two categories of represented spaces, one equipped with the computable maps, and one equipped with the continuous maps. We call the resulting structure a \emph{category extension} (cf.~\cite{paulysearchproblems,paulysearchproblemscie}), as the former is a subcategory of the latter, and shares its structure (products, coproducts, exponentials). In general, all results on represented spaces have two instances, one for the computable and one for the continuous maps, with the proofs being identical. Essentially, the continuous case can be considered as the relativization of the computable one, as a function on Baire space is continuous iff it is computable relative to some oracle. The presence of two categories with the same objects and the same structure was already observed by \name{Bauer} (in a slightly different setting) in 1998 \cite{bauer5}.

Given two represented spaces $\mathbf{X}$, $\mathbf{Y}$ we obtain a third represented space $\mathcal{C}(\mathbf{X}, \mathbf{Y})$ of functions from $X$ to $Y$ by letting $0^n1p$ be a $[\delta_X \to \delta_Y]$-name for $f$, if the $n$-th Turing machine equipped with the oracle $p$ computes a realizer for $f$. As a consequence of the UTM theorem, $\mathcal{C}(-, -)$ is the exponential in the category of continuous maps between represented spaces, and the evaluation map is even computable (as are the other canonic maps, e.g.~currying).

This function space constructor, together with two represented spaces , $\mathbb{N} = (\mathbb{N}, \delta_\mathbb{N})$ and $\mathbb{S} = (\{\bot, \top\}, \delta_\mathbb{S})$, allows us to obtain a model of \name{Escard\'o}'s synthetic topology \cite{escardo}. The representation are given by $\delta_\mathbb{N}(0^n10^\mathbb{N}) = n$, $\delta_\mathbb{S}(0^\mathbb{N}) = \bot$ and $\delta_\mathbb{S}(p) = \top$ for $p \neq 0^\mathbb{N}$. It is straightforward to verify that the computability notion for the represented space $\mathbb{N}$ coincides with classical computability over the natural numbers. The \Sierp space $\mathbb{S}$ in turn allows us to formalize semi-decidability. The computable functions $f : \mathbb{N} \to \mathbb{S}$ are exactly those where $f^{-1}(\{\top\})$ is recursively enumerable (and thus $f^{-1}(\{\bot\})$ co-recursively enumerable).

In general, for any represented space $\mathbf{X}$ we obtain two spaces of subsets of $\mathbf{X}$; the space of open sets $\mathcal{O}(\mathbf{X})$ by identifying $f \in \mathcal{C}(\mathbf{X}, \mathbb{S})$ with $f^{-1}(\{\top\})$, and the space of closed sets $\mathcal{A}(\mathbf{X})$ by identifying $f \in \mathcal{C}(\mathbf{X}, \mathbb{S})$ with $f^{-1}(\{\bot\})$. The properties of the spaces of open and closed sets, namely computability of the usual operations, follow from a few particular computable functions on \Sierp space $\mathbb{S}$ and the fundamental function space properties.

\begin{proposition}
\label{prop:sierpbasics}
The functions $\wedge, \vee : \mathbb{S} \times \mathbb{S} \to \mathbb{S}$ and $\bigvee : \mathcal{C}(\mathbb{N}, \mathbb{S}) \to \mathbb{S}$ are computable.
\end{proposition}

One useful consequence of staying within the category when forming the space $\mathcal{O}(\mathbf{X})$ is that we can iterate this to obtain $\mathcal{O}(\mathcal{O}(\mathbf{X}))$, a space appearing in several further constructions. We introduce the space $\mathcal{K}(\mathbf{X})$ of compact sets by identifying a set $A \subseteq \mathbf{X}$ with $\{U \in \mathcal{O}(\mathbf{X}) \mid A \subseteq U\} \in \mathcal{O}(\mathcal{O}(\mathbf{X})$. To ensure well-definedness, we restrict the sets $A$ to saturated sets. As a dual notion, we find the space of overt set $\mathcal{V}(\mathbf{X})$ by identifying $A \subseteq \mathbf{X}$ with $\{U \in \mathcal{O}(\mathbf{X}) \mid A \cap U \neq \emptyset\} \in \mathcal{O}(\mathcal{O}(\mathbf{X})$. The canonization operation here is the topological closure.

There always is a canonic computable map $\kappa_\mathbf{X} : \mathbf{X} \to \mathcal{O}(\mathcal{O}(\mathbf{X}))$ defined via $\kappa_\mathbf{X}(x) = \{U \mid x \in U\}$. Using the spaces introduced above, we can read $\kappa_\mathbf{X} : \mathbf{X} \to \mathcal{K}(\mathbf{X})$ as $\kappa_\mathbf{X}(x) = \{x\}\uparrow$ or $\kappa_\mathbf{X} : \mathbf{X} \to \mathcal{V}(\mathbf{X})$ as $\kappa_\mathbf{X}(\{x\}) = \overline{\{x\}}$ instead. The image of $\mathbf{X}$ under $\kappa_\mathbf{X}$ shall be denoted by $\mathbf{X}_\kappa$. The following definition essentially goes back to Schr\"oder \cite{schroder5} and provides an effective counterpart to the definition in \cite{schroder}:

\begin{definition}
A space $\mathbf{X}$ is called \emph{computably admissible}, if $\mathbf{X}$ and $\mathbf{X}_\kappa$ are computably isomorphic.
\end{definition}

Note that $\mathbf{X}_\kappa$ is always computably admissible, i.e.~isomorphic to $(\mathbf{X}_\kappa)_\kappa$. The computably admissible spaces are precisely those that can be regarded as topological spaces, based on the fact that the computable map $f \mapsto f^{-1} : \mathcal{C}(\mathbf{X}, \mathbf{Y}) \to \mathcal{C}(\mathcal{O}(\mathbf{Y}), \mathcal{O}(\mathbf{X}))$ becomes computably invertible iff $\mathbf{Y}$ is computably admissible.

Some further properties of spaces we will require are being \emph{computably Hausdorff}, iff $x \mapsto \{x\} : \mathbf{X} \to \mathcal{A}(\mathbf{X})$ is computable, being \emph{computably discrete} iff $x \mapsto \{x\} : \mathbf{X} \to \mathcal{O}(\mathbf{X})$ is computable, being \emph{computably compact} iff $\operatorname{IsEmpty} : \mathcal{A}(\mathbf{X}) \to \mathbb{S}$ is computable, and being \emph{computably overt} iff $\operatorname{IsNonEmpty} : \mathcal{O}(\mathbf{X}) \to \mathbb{S}$ is computable.

To see that we may regard Polish spaces as a special case of represented spaces, we define computable metric spaces following \name{Weihrauch}'s \cite{weihrauchi}. The computable Polish spaces, are derived from complete computable metric spaces by forgetting the details of the metric, and just retaining the representation (or rather, the equivalence class of representations under computable translations). Forgetting the metric is relevant when it comes to compatibility with definitions in effective descriptive set theory as shown in \cite{pauly-gregoriades-arxiv} by \name{Gregoriades}, \name{Kisp\'eter} and the author. To be precise, a fundamental notion used in \cite{moschovakis,gregoriades} is a recursively presented metric space, and this is a more restrictive notion than a computable metric space. Rescaling the metric however is sufficient to overcome the distinction, thus the notions of a computable Polish space and of an effective Polish space do coincide.

\begin{definition}
\label{def:cms}
We define a computable metric space with its Cauchy representation such that:
\begin{enumerate}
\item A computable metric space is a tuple $\mathbf{M} = (M, d, (a_n)_{n \in \mathbb{N}})$ such that $(M,d)$ is a metric space and $(a_n)_{n \in\mathbb{N}}$ is a dense sequence in $(M,d)$.
\item The relation
\[ \{(t,u,v,w) \: |\: \nu_{\mathbb{Q}}(t) < d(a_u, a_v) <\nu_{\mathbb{Q}}(w) \} \text{ is recursively enumerable.} \]
\item The Cauchy representation $ \delta_{\mathbf{M}} \: : \: \Baire \rightharpoonup M $ associated with the computable metric space $ \mathbf{M} =  (M, d, A, \alpha) $ is defined by
\[ \delta_{\mathbf{M}}(p) = x \: : \: \Longleftrightarrow  \begin{cases}
      d(a_{p(i)}, a_{p(k)}) \leq 2^{-i} \text{ for } i < k\\
   \text{and } x = \lim\limits_{i\rightarrow \infty}a_{p(i)}
  \end{cases} \]

\end{enumerate}
\end{definition}
\section{Synthetic descriptive set theory}
\label{section:synthetic}
We will now see how to \emph{lift} the notions from synthetic topology (instantiated with represented spaces) to obtain the relevant concepts for descriptive set theory. The presentation is taken from \cite[Section 2]{pauly-descriptive} by the author and \name{de Brecht}, where also the missing proofs can be found. The fundamental new notion for synthetic descriptive set theory are endofunctors on the category of continuous functions between represented spaces. An endofunctor is an operation $d$ from a category to itself which maps objects to objects, morphisms to morphisms, preserves identity morphisms,
and is compatible with composition, i.e.~$d(f \circ g) = (df) \circ (dg)$. For any two represented spaces $\mathbf{X}$, $\mathbf{Y}$, an endofunctor $d$ induces\footnote{In the presence of exponentials and a final object, an endofunctor $d$ may have an internal characterization. For fixed objects $\mathbf{X}$, $\mathbf{Y}$, let $D : \mathcal{C}(\mathbf{X}, \mathbf{Y}) \to \mathcal{C}(d\mathbf{X}, d\mathbf{Y})$ be an internal realization of $d$, if the following holds: Let $f : \mathbf{X} \to \mathbf{Y}$ be a morphism, and $f' : \mathbf{X} \times \mathbf{1} \to \mathbf{Y}$ the corresponding morphism up to equivalence. By definition of the exponential, we then have a map $\lambda f' : \mathbf{1} \to \mathcal{C}(\mathbf{X}, \mathbf{Y})$. In the same way, there is a map $\lambda (df)' : \mathbf{1} \to \mathcal{C}(d\mathbf{X}, d\mathbf{Y})$. The criterion now is $\lambda (df)' = D \circ \lambda f'$.} a map $d : \mathcal{C}(\mathbf{X}, \mathbf{Y}) \to \mathcal{C}(d\mathbf{X}, d\mathbf{Y})$. If this map is always computable, we call $d$ computable\footnote{Modulo the continuity/computability distinction, this would be a special case of an enriched endofunctor, if we understand a cartesian closed category to be enriched over itself.}. In the following, $d$ shall always be some computable endofunctor.

The typical examples relevant for our development of descriptive set theory will be operators that keep the underlying set of a represented spaces the same, and modify the representation in a sufficiently uniform way to ensure the requirements for computable endofunctors. Such operators have been called \emph{jump operators} in \cite{debrecht5}, and specific examples can be found both there, in \cite[Section 3]{pauly-descriptive} and below. For computable endofunctors that do change the underlying sets in a significant way, the interpretation of many of the following definitions becomes less clear, but an example of a computable endofunctor that still produces sensible notions is given in \cite[Subsection 3.5]{pauly-descriptive}.

The computable endofunctors we study correspond to classes of sets such as $\Sigma_2^0$, $\Sigma_3^0$, $\Delta_2^0$, etc.; with the closure properties of the set-classes depending on how the endofunctor interacts with products. We say that $d$ preserves binary products, if $d(\mathbf{X} \times \mathbf{X}) \cong d\mathbf{X} \times d\mathbf{X}$ (where $\cong$ denotes computable isomorphism) for any represented space $\mathbf{X}$, and that $d$ preserves products if $d\mathcal{C}(\mathbf{N}, \mathbf{X}) \cong \mathcal{C}(\mathbb{N}, d\mathbf{X})$ for any represented space $\mathbf{X}$.

\subsection{The $d$-open sets}
For a represented space $\mathbf{X}$, we shall call $\mathcal{C}(\mathbf{X}, d\mathbb{S})$ the space of $d$-open sets $\mathcal{O}^d(\mathbf{X})$. If $d\mathbf{S}$ still has the underlying set $\{\bot, \top\}$ the elements of $\mathcal{O}^d(\mathbf{X})$ actually are subsets of $X$ in the usual way\footnote{Which is to identify a function $f : X \to \{\bot, \top\}$ with the set $f^{-1}(\top)$, and vice versa a set with its characteristic function.}, otherwise this is a purely abstract definition. The complements of $d$-open sets are $d$-closed sets, denoted by $\mathcal{A}^d(\mathbf{X})$. A variety of nice closure properties follows immediately, with the proofs being straight-forward modifications of those for the corresponding results for open sets in \cite[Proposition 6]{pauly-synthetic-arxiv}:

\begin{proposition}
\label{prop:dopensets}
The following operations are computable for any represented spaces $\mathbf{X}, \mathbf{Y}$:
\begin{enumerate}
\item $(f, U) \mapsto f^{-1}(U) : \mathcal{C}(\mathbf{X}, \mathbf{Y}) \times \mathcal{O}^d(\mathbf{Y}) \to \mathcal{O}^d(\mathbf{X})$
\item $\operatorname{Cut} : \mathbf{Y} \times \mathcal{O}^d(\mathbf{X} \times \mathbf{Y}) \to \mathcal{O}^d(\mathbf{X})$ mapping $(y, U)$ to $\{x \mid (x, y) \in U\}$
    \seti
\end{enumerate}
If $d$ preserves binary products, we additionally obtain:
\begin{enumerate}
\conti
\item $\cap, \cup : \mathcal{O}^d(\mathbf{X}) \times \mathcal{O}^d(\mathbf{X}) \to \mathcal{O}^d(\mathbf{X})$
\item $\times : \mathcal{O}^d(\mathbf{X}) \times \mathcal{O}^d(\mathbf{Y}) \to \mathcal{O}^d(\mathbf{X} \times \mathbf{Y})$
\seti
\end{enumerate}
If $d$ preserves products, we additionally obtain:
\begin{enumerate}
\conti
\item $\bigcup : \mathcal{C}(\mathbb{N}, \mathcal{O}^d(\mathbf{X})) \to \mathcal{O}^d(\mathbf{X})$
\end{enumerate}
\end{proposition}

\subsection{$d$-continuity and $d$-measurability}
Now we can introduce the notion of $d$-measurability: We call a function $f : \mathbf{X} \to \mathbf{Y}$ $d$-measurable, if $f^{-1} : \mathcal{O}(\mathbf{Y}) \to \mathcal{O}^d(\mathbf{X})$ is well-defined and continuous, i.e.~if the preimages of open sets under $f$ are uniformly $d$-open. The $d$-measurable functions from $\mathbf{X}$ to $\mathbf{Y}$ thus form a represented space $\mathcal{C}^d(\mathbf{X}, \mathbf{Y})$, which is by construction homeomorphic to a subspace of $\mathcal{C}(\mathcal{O}(\mathbf{Y}), \mathcal{O}^d(\mathbf{X}))$.

A $d$-continuous function from $\mathbf{X}$ to $\mathbf{Y}$ shall just be a continuous function $f : \mathbf{X} \to d\mathbf{Y}$. Note again, that if $d$ alters the underlying sets, then a $d$-continuous function between represented spaces will not necessarily induce a function on the underlying sets. The notion of $d$-continuity is a generalization of the Kleisli-morphisms w.r.t.~a monad -- if $d$ can be turned into a monad, then the $d$-continuous functions are precisely the Kleisli-morphisms. Some, but not all, of our examples of computable endofunctors will actually be monads in a natural way.

Both the $d$-measurable and the $d$-continuous functions have some of the closure properties expected from classes of $\mathfrak{B}$-measurable functions. For their formulation, note that the composition of two computable endofunctors is a computable endofunctor again.

\begin{proposition}
\label{prop:dcontdmeas}
Both $d$-continuous and $d$-measurable maps are closed under composition with continuous maps from both sides, i.e.~the following maps are computable for any represented spaces $\mathbf{X}$, $\mathbf{Y}$, $\mathbf{Z}$:
\begin{enumerate}
\item $\circ : \mathcal{C}(\mathbf{X}, \mathbf{Y}) \times \mathcal{C}(\mathbf{Y}, d\mathbf{Z}) \to \mathcal{C}(\mathbf{X}, d\mathbf{Z})$
\item $\circ : \mathcal{C}(\mathbf{X}, \mathbf{Y}) \times \mathcal{C}^d(\mathbf{Y}, \mathbf{Z}) \to \mathcal{C}^d(\mathbf{X}, \mathbf{Z})$
\item $\circ : \mathcal{C}(\mathbf{X}, d\mathbf{Y}) \times \mathcal{C}(\mathbf{Y}, \mathbf{Z}) \to \mathcal{C}(\mathbf{X}, d\mathbf{Z})$
\item $\circ : \mathcal{C}^d(\mathbf{X}, \mathbf{Y}) \times \mathcal{C}(\mathbf{Y}, \mathbf{Z}) \to \mathcal{C}^d(\mathbf{X}, \mathbf{Z})$
\seti
\end{enumerate}
More generally, we can consider a second computable endofunctor $e$ and obtain:
\begin{enumerate}
\conti
\item $\circ : \mathcal{C}(\mathbf{X}, e\mathbf{Y}) \times \mathcal{C}(\mathbf{Y}, d\mathbf{Z}) \to \mathcal{C}(\mathbf{X}, ed\mathbf{Z})$
\seti
\end{enumerate}
Taking into consideration the definition of $\mathcal{O}^d(\mathbf{X})$ as $\mathcal{C}(\mathbf{X}, d\mathbb{S})$, we get the special case:
\begin{enumerate}
\conti
\item $(f, U) : \mathcal{C}(\mathbf{X}, e\mathbf{Y}) \times \mathcal{O}^d(\mathbf{Y}) \to \mathcal{O}^{ed}(\mathbf{X})$
\item $\circ : \mathcal{C}(\mathbf{X}, e\mathbf{Y}) \times \mathcal{C}^d(\mathbf{Y}, \mathbf{Z}) \to \mathcal{C}^{ed}(\mathbf{X}, \mathbf{Z})$
\seti
\end{enumerate}
Finally, we find that $e$-continuity uniformly implies $e$-measurability:
\begin{enumerate}
\conti
\item $\id : \mathcal{C}(\mathbf{X}, e\mathbf{Y}) \to \mathcal{C}^e(\mathbf{X}, \mathbf{Y})$
\end{enumerate}
\end{proposition}

\subsection{$d$-admissibility}
Having seen that $d$-continuity always implies $d$-measurability, we now strive for conditions that make the converse implication true, as well. Noting that $\id$-continuity is continuity of maps between represented spaces, and $\id$-measurability (uniform) topological continuity, we see that we need a notion of $d$-admissibility.

As a special case of Proposition \ref{prop:dcontdmeas} (8) with $\mathbf{X} = \mathbf{1}$ and using trivial isomorphisms, we obtain the computability of a canonic mapping $\kappa^d : d\mathbf{Y} \to \mathcal{C}(\mathcal{O}(\mathbf{Y}), d\mathbb{S})$. The image of $d\mathbf{Y}$ under $\kappa^d$ shall be denoted by $\kappa^d\mathbf{Y}$ (not by $\kappa^dd\mathbf{Y}$!!).

\begin{proposition}
\label{prop-admissibility}
The following are equivalent:
\begin{enumerate}
\item $\id : \mathcal{C}(\mathbf{X}, d\mathbf{Y}) \to \mathcal{C}^d(\mathbf{X}, \mathbf{Y})$ is computably invertible for any represented space $\mathbf{X}$.
\item $\kappa^d : d\mathbf{Y} \to \mathcal{C}(\mathcal{O}(\mathbf{Y}), d\mathbb{S})$ is computably invertible.
\item $\kappa^d\mathbf{Y} \cong d\mathbf{Y}$.
\end{enumerate}
\end{proposition}

A space $\mathbf{Y}$ satisfying these equivalent conditions shall be called $d$-admissible. We observe the following:

\begin{proposition}
\label{sadmissible}
$\mathbb{S}$ is $d$-admissible.
\end{proposition}

Now, consider $\kappa^d$ as an operation on the whole category of continuous functions between represented spaces. It is not hard to verify that $\kappa^d$ itself is a computable endofunctor. Even more, we can consider $d \mapsto \kappa^d$ ($=: \kappa$) as an operation on computable endofunctors! As a consequence of Proposition \ref{sadmissible} we obtain:

\begin{corollary}
$\kappa^{(\kappa^d)} \cong \kappa^d$.
\end{corollary}

\begin{corollary}
Every represented space is $\kappa^d$-admissible.
\end{corollary}

\begin{corollary}
$\mathcal{O}^d(\mathbf{X}) = \mathcal{O}^{\kappa^d}(\mathbf{X})$.
\end{corollary}

\begin{corollary}
$d$-measurability and $\kappa^d$-continuity coincide.
\end{corollary}

\begin{corollary}
If $\mathbf{Y}$ is $d$-admissible, then $\mathcal{C}^d(\mathbf{X}, \mathbf{Y})$ and $\mathcal{C}(\mathbf{X}, d\mathbf{Y})$ are homeomorphic.
\end{corollary}

\begin{corollary}
If $\mathbf{Y}$ is $e$-admissible, then $\circ : \mathcal{C}^e(\mathbf{X}, \mathbf{Y}) \times \mathcal{C}^d(\mathbf{Y}, \mathbf{Z}) \to \mathcal{C}^{ed}(\mathbf{X}, \mathbf{Z})$ is computable.
\end{corollary}

For a large class of spaces and computable endofunctors, we can provide admissibility results without having to resort to modifying the endofunctor. We start with the seemingly innocuous:
\begin{proposition}
Let $d$ preserve products. Then $\mathcal{O}(\mathbb{N})$ is $d$-admissible.
\end{proposition}

\begin{corollary}
\label{corr:productsadmissibility}
Let $d$ preserve products, and let $\mathbf{X}$ be countably based and admissible. Then $\mathbf{X}$ is $d$-admissible.
\end{corollary}

The preceding corollary relies on \name{Weihrauch}'s observation \cite{weihrauchd} that the countably-based admissible spaces are just the subspaces of $\mathcal{O}(\mathbb{N})$, together with $d$-admissibility being closed under formation of subspaces. Additionally, it may be the reason that countably-based $T_0$-spaces seem to form a natural demarkation line for the extension of descriptive set theory \cite{debrecht6}. Combining its statement with Proposition \ref{prop:dopensets}, we see that any computable endofunctor preserving products nicely characterizes a $\Sigma$-like class of sets and the corresponding measurable functions on all countably based admissible spaces.

\subsection{Further concepts}
The other concepts from synthetic topology studied for represented spaces in \cite{pauly-synthetic-arxiv}, namely Hausdorff, discreteness, compactness and overtness, can also be lifted along some endofunctor, and retain most of their nice properties. Rather than listing all of these statements and definitions, we shall only consider those used later in applications.

\begin{definition}
A space $\mathbf{X}$ is called computably $d$-Hausdorff, iff $x \mapsto \{x\} : \mathbf{X} \to \mathcal{A}^d(\mathbf{X})$ is computable.
\end{definition}

\begin{proposition}
The following are equivalent:
\begin{enumerate}
\item $\mathbf{X}$ is computably $d$-Hausdorff.
\item $\mathalpha{\neq} : \mathbf{X} \times \mathbf{X} \to d\mathbb{S}$ is computable.
\seti
\end{enumerate}
If $d$ preserves binary products, then the following are also equivalent to those above:
\begin{enumerate}
\conti
\item $\{(x,x) \mid x \in \mathbf{X}\} \in \mathcal{A}^d(\mathbf{X} \times \mathbf{X})$ is computable.
\item $\operatorname{Graph} : \mathcal{C}(\mathbf{Y}, \mathbf{X}) \to \mathcal{A}^d(\mathbf{Y} \times \mathbf{X})$ is well-defined and computable for any represented space $\mathbf{Y}$.
\end{enumerate}
\end{proposition}

Some properties related to $d$-Hausdorff have been studied by \name{Schr\"oder} and \name{Selivanov} in \cite{selivanov4,selivanov4b}.

\begin{proposition}
If $d$ preserves binary products and $\mathbf{X}$ is computably Hausdorff, then $d\mathbf{X}$ is computably $d$-Hausdorff.
\end{proposition}

\begin{definition}
A space $\mathbf{X}$ is called $d$-overt, iff $\operatorname{IsNonEmpty} : \mathcal{O}^d(\mathbf{X}) \to d\mathbb{S}$ is computable.
\end{definition}

\subsection{The Markov-variant}
In effective descriptive set theory, the notion of affectivity between higher-order objects being employed often is not computability, but rather Markov computability. A function $f : \mathbf{X} \to \mathbf{Y}$ is called Markov-computable, if there is some computable partial  function $\phi : \subseteq \mathbb{N} \to \mathbb{N}$, such that whenever $i$ is an index of a computable element in $\mathbf{X}$, then $\phi(i)$ is an index of $f(i)$. Any computable function is Markov-computable, while the converse fails.

Subsequently, an endofunctor $d$ is called Markov-computable, if any $d : \mathcal{C}(\mathbf{X}, \mathbf{Y}) \to \mathcal{C}(d\mathbf{X}, d\mathbf{Y})$ is Markov-computable\footnote{Note that just as a computable endofunctor is linked to the topological jump operators of \cite{debrecht5}, Markov-computable endofunctors are linked to the computable jump operators.}. The effective measurability notion going with Markov-computable endofunctors is (weak) non-uniform computability, i.e.~if for any computable $U \in \mathcal{O}(\mathbf{Y})$ we find $f^{-1}(U) \in \mathcal{O}^d(\mathbf{X})$ to be computable, we call $f$ to be Markov-$d$-measurable. The represented space $\mathcal{C}^{\mathcal{M}d}(\mathbf{X}, \mathbf{Y})$ of Markov-$d$-measurable functions essentially represents a function by some oracle $p$ paired with a table listing indices of computably open sets and their $p$-computably $d$-open preimages.

\begin{proposition}
Let $d$ be Markov-computable. Then $\id : \mathcal{C}(\mathbf{X}, d\mathbf{Y}) \to \mathcal{C}^{\mathcal{M}d}(\mathbf{X}, \mathbf{Y})$ is computable.
\end{proposition}

We can define the Markov-variant of $\kappa^d$ via letting $\eta^d : d\mathbf{Y} \to \mathcal{C}^{\mathcal{M}d}(\mathbf{1}, \mathbf{Y})$ be the canonic map, and subsequently obtain a notion of Markov-$d$-admissibility with just the same properties as before.

\subsection{Adjoint endofunctors}
\label{subsec:adjointness}
A computable endofunctor $d$ is computably-left-adjoint to a computable endofunctor $e$ (and $e$ is right-adjoint to $d$), if $\mathcal{C}(d\mathbf{X},\mathbf{Y})$ and $\mathcal{C}(\mathbf{X},e\mathbf{Y})$ are computably isomorphic, and the isomorphisms are natural in $\mathbf{X}$ and $\mathbf{Y}$.

Likewise, a Markov-computable endofunctor $d$ is Markov-computably-left-adjoint to a Markov-computable endofunctor $d$, if $\mathcal{C}(d\mathbf{X},\mathbf{Y})$ and $\mathcal{C}(\mathbf{X},e\mathbf{Y})$ are Markov-computably isomorphic, and the isomorphisms are natural in $\mathbf{X}$ and $\mathbf{Y}$. Note that a statement that $\mathbf{X}$ and $\mathbf{Y}$ are Markov-computably isomorphic only refers to the cardinality (as there has to be a bijection) and to the computable elements. Note further that for computable endofunctors being Markov-computably-adjoint is a weaker condition than being computably adjoint (and that both concepts formally make sense).

At the current state, we do not have interesting examples of pairs of computably-adjoint computable endofunctors. We will discuss two cases of Markov-computably adjoint Markov-computable functors later.

It is quite illuminating to see the special case of the definitions above where $\mathbf{Y} := \mathbb{S}$. We see that if $d$ is (Markov)-computably-left-adjoint to $e$, then the (computably) $e$-open subsets of $\mathbf{X}$ are precisely the (computably) open subsets of $d\mathbf{X}$. This aspect of our two examples below has been utilized before.

It is a central fact in the study of pairs of adjoint functors in category theory that their composition induces a monad. Some consequences of this of interest for our theory are the following:
\begin{proposition}
\label{prop:adjoint}
Let the (Markov)-computable endofunctor $d$ be (Markov)-computably-left-adjoint to the (Markov)-computable endofunctor $e$. Then:
\begin{enumerate}
\item There is a canonic computable\footnote{$\eta_\mathbf{X}$ is indeed always computable, even if the endofunctors involved are only Markov-computable. The same pattern applies in the following.} unit map $\eta_\mathbf{X} : \mathbf{X} \to ed\mathbf{X}$.
\item $e\mathbf{Y}$ and $ede\mathbf{Y}$ are computably isomorphic.
\item $\mathalpha{\circ} : \mathcal{C}(\mathbf{X},ed\mathbf{Y}) \times \mathcal{C}(\mathbf{Y},e\mathbf{Z}) \to \mathcal{C}(\mathbf{X},e\mathbf{Z})$ is well-defined and computable.
\item $(de) \cong (de)(de)$.
\end{enumerate}
\end{proposition}

Items $(3.) \& (4.)$ in the preceding proposition shows that if, given some endofunctor $e$, we can find a Markov-computably-left-adjoint $d$ for it, then we obtain a class functions (namely the $de$-continuous ones) that is closed under composition, and that if composed with an $e$-continuous function from the right, again yield an $e$-continuous function. This seems to be the concept underlying the study of lowness notions in computability theory (e.g.~\cite{nies}): A recursion theoretic notion induces a Markov-computable endofunctor, and the corresponding lowness notion then is the monad obtained from a suitable adjoint endofunctor.

The importance of adjointness had already been noticed \name{de Brecht} in \cite{debrecht5}.

\section{The Borel-hierarchy and the $'$-operator}
\label{sec:borel}
A central part of descriptive set theory is the Borel hierarchy. As a testimonial to its naturality, we will see how the underlying concept (as revealed by studying it in the context of represented spaces) shows up in a variety of settings. This includes the known close connection to the arithmetical hierarchy in recursion theory.

For now, consider a Polish space $\mathbf{X}$. Now let $\Sigma_1^0(\mathbf{X}) := \mathcal{O}(\mathbf{X})$, $\Pi_\alpha^0(\mathbf{X}) := \{X \setminus U \mid U \in \Sigma_\alpha^0(\mathbf{X})\}$, $\Sigma_{\alpha+1}^0(\mathbf{X}) = \{\bigcup_{i \in \mathbb{N}} A_i \mid \forall i \in \mathbb{N} \ A_i \in \Pi_\alpha^0(\mathbf{X})\}$ and $\Sigma_{\beta}^0(\mathbf{X}) = \bigcup_{\alpha < \beta} \Sigma_{\alpha}^0(\mathbf{X})$ for limit ordinals $\beta$. Moreover, let $\Delta_\alpha^0(\mathbf{X}) = \Sigma_\alpha^0(\mathbf{X}) \cap \Pi_\alpha^0(\mathbf{X})$.

The $\Sigma_\alpha^0$-sets behave in some ways like the open set: They are closed under countable unions and finite intersections, and the preimages of a $\Sigma_\alpha^0$-set under a continuous function is a $\Sigma_\alpha^0$-set again. We also find that $\Sigma_\alpha^0(\mathbf{X}) \subseteq \Sigma_{\alpha'}^0(\mathbf{X})$ if $\alpha < \alpha'$.

For non-metric topological spaces that are still countably based and $T_0$, \name{Selivanov} \cite{selivanov3} suggest a modified definition of the Borel hierarchy, using $\Sigma_{\alpha+1}^0(\mathbf{X}) := \{\bigcup_{i \in \mathbb{N}} (U_i \setminus U_i') \mid \forall i \in \mathbb{N} \ U_i, U'_i \in \Sigma_\alpha^0(\mathbf{X})\}$ instead. This modification ensures that $\Sigma_\alpha^0(\mathbf{X}) \subseteq \Sigma_{\alpha'}^0(\mathbf{X})$ if $\alpha < \alpha'$ remains true, and is equivalent to the original definitions for metric spaces.

To see how to obtain the Borel hierarchy from a computable endofunctor, consider the map $\lim : \subseteq \Baire \to \Baire$ defined via $\lim(p)(n) = \lim_{i \to \infty} p(\langle n, i\rangle)$. This map was studied in the context of Weihrauch reducibility by \name{von Stein} \cite{stein}, \name{Mylatz} \cite{Mylatz} and \name{Brattka} \cite{brattka11,brattka,brattka4}, with the latter noting in \cite{brattka} the connections to the Borel hierarchy. It also appears in the context of model of hypercomputation as shown by \name{Ziegler} \cite{ziegler2,ziegler7}, and captures precisely the additional computational power certain solutions to general relativity could provide beyond computability \cite{hogarth}.

Now $\lim$ induces a computable endofunctor $'$ via $(X, \delta_X)' = (X, \delta_X \circ \lim)$ and $(f : \mathbf{X} \to \mathbf{Y})' = f : \mathbf{X}' \to \mathbf{Y}'$. We iterate this endofunctor, so let $\mathbf{X}^{(0)} = \mathbf{X}$, $\mathbf{X}^{(\alpha+1)} = (\mathbf{X}^{(\alpha)})'$ and $\mathbf{X}^{(\beta)} = \pi_2(\coprod_{\gamma < \beta} \mathbf{X}^{(\gamma)})$ for limit\footnote{The definition for limit ordinals was suggested by \name{Bauer} at CCA 2009.} ordinals $\beta$.

\begin{proposition}[{\cite[Proposition 23]{pauly-descriptive}}]
\label{prop:jumpproducts}
$^{(\alpha)}$ is a computable endofunctor preserving binary products. Moreover, $^{(\alpha+1)}$ even preserves products.
\end{proposition}

Besides a slight misalignment regarding the numbers, the $^{(\alpha)}$-open sets correspond to closely to the $\Sigma_\alpha^0$-sets. Given some countable ordinal $\alpha > 0$, let $\overline{\alpha} = \alpha - 1$ for $\alpha < \omega_0$ and $\overline{\alpha} = \alpha + 1$ otherwise.

\begin{proposition}[\cite{debrecht4}, based on \cite{saintraymond}]
\label{prop:sigmajump}
For a quasi-Polish space $\mathbf{X}$, the elements of $\mathcal{O}^{(\overline{\alpha})}(\mathbf{X})$ are precisely the $\Sigma_{\alpha}^0$-sets in Selivanov's definition.
\end{proposition}

It was shown by \name{Gregoriades}, \name{Kisp\'eter} and the author in \cite[Section 5]{pauly-gregoriades-arxiv} that the continuity structure the space $\mathcal{O}^{(\alpha)}(\mathbf{X})$ carries, does correspond precisely to the structure given by a good universal system used in effective descriptive set theory (for Polish spaces).

Now combining Proposition \ref{prop:jumpproducts} with Corollary \ref{corr:productsadmissibility} gives us:
\begin{corollary}
Any admissible countably based space is $^{(\alpha+1)}$-admissible.
\end{corollary}

To spell it out, if $\mathbf{X}$, $\mathbf{Y}$ are admissible and countably based, then a function $f : \mathbf{X} \to \mathbf{Y}$ is $\Sigma_\alpha^0$-measurable iff $f : \mathbf{X} \to \mathbf{Y}^{(\overline{\alpha})}$ is continuous. This can be understood as a generalization of the Lebesgue-Hausdorff-Banach theorem for Polish spaces: Say that the Baire class 0 functions are the continuous functions, the Baire class $1$ functions the $\Sigma_2^0$-measurable functions\footnote{For some special metric spaces, the following theorem would hold without explicitly demanding truth for $\alpha = 1$, i.e.~with the Baire class 1 functions being the point-wise limits of continuous functions. This fails for other spaces, though: If $\mathbf{X}$ is connected and $\mathbf{Y}$ discrete, then point-wise limits of continuous functions are continuous themselves (Example taken from \cite{mottoros4}). Such exceptions marring the theory disappear when moving to the synthetic approach.}, the Baire class $\alpha$ functions the point-wise limits functions of Baire class $< \alpha$. Now we can formulate the:

\begin{theorem}[Lebesgue -- Hausdorff -- Banach]
\label{theo:lhbt}
Let $\mathbf{X}$, $\mathbf{Y}$ be separable metric spaces. Then a function $f : \mathbf{X} \to \mathbf{Y}$ is Baire class $\alpha$ iff it is $\Sigma_{\alpha+1}^0$-measurable.
\end{theorem}

In Subsection \ref{subsec:adjointness} the role of adjointness was mentioned, including its connections to lowness notions in computability theory. Now $'$ has a Markov-computably left adjoint, which we shall briefly consider. Let $J : \Cantor \to \Cantor$ be the Turing-jump (i.e. $J(p)$ is the Halting problem relative to $p$), and then define $\int$ via $\int(X, \delta_X) = (X, \delta_X \circ J^{-1})$ with the straight-forward extension to morphisms. This yields a Markov-computable endofunctor.

As observed in more general terms in Subsection \ref{subsec:adjointness}, the computably open subsets of $\int \mathbf{X}$ are just the computably $\Sigma^0_2$-subsets of $\mathbf{X}$. Under this perspective, the space $\int \Cantor$ had already been investigated in \cite{miller}.

Note that $J^{-1}$ is computable, whereas $J$ is not, hence $\id : \int \mathbf{X} \to \mathbf{X}$ is computable, and $\id : \mathbf{X} \to \int \mathbf{X}$ typically not.

Now the low-endofunctor $^\vee$ is defined via $\mathbf{X}^\vee = (\int \mathbf{X})'$. Both $\int$ and $\vee$ were studied in \cite{paulybrattka,brattka4} by \name{Brattka}, \name{de Brecht} and the author. The results there are essentially special cases of Proposition \ref{prop:adjoint}: In particular, we see that $(^\vee)(^\vee) \cong ^\vee$, and that if $f$ is $'$-continuous and $g$ is $^\vee$-continuous, then $f \circ g$ is $'$-continuous again (hence the name \emph{low}). Note that $p \in \Cantor$ is low in the traditional sense, iff $p \in \left (\Cantor \right )^\vee$ is computable.

\begin{theorem}[{\cite[Theorem 36]{pauly-descriptive}}]
A Polish space is $'$-overt iff it is $K_\sigma$.
\end{theorem}

\section{Point degree spectra of represented spaces}
\label{sec:pointspectra}
The concept of a \emph{point degree spectrum} links the study of degree structures in computability theory to the study of dimension in topology. They were introduced by \name{Kihara} and the author in \cite{pauly-kihara-arxiv}, from where much of this section's material is taken.

In a broader context, there are various instances of properties of spaces and sets that start making sense for points in an effective treatment. Martin-L\"of randomness of points corresponds to Lebesgue measure of sets and 1-genericity is the analogue to being co-meagre \cite{nies}. Likewise, effective Hausdorff dimension \cite{lutz,lutz2,lutz3} is applicable to individual points. In all these cases, individual points can carry some amount of complexity -- e.g.~a Martin-L\"of random point is in some sense too complicated to be included in computable $\Pi^0_2$-set having effectively measure $0$. Our work provides an effective notion corresponding to topological invariants such as small inductive dimension or metrizability, and e.g.~allows us to say that certain points are too complicated to be (computably) a member of a (finite-dimensional) Polish space.

A cornerstone of the framework of point degree spectra are the Medvedev degrees $\mathfrak{M}$ \cite{medvedev}. These are obtained by taking equivalence classes from Medvedev reducibility $\leq_M$, defined on subsets $A$, $B$ of Baire space $\Baire$ via $A \leq_M B$ iff there is a computable function $F : B \to A$. Important substructures of $\mathfrak{M}$ also relevant to us are the Turing degrees $\mathfrak{T}$, the continuous degrees $\mathfrak{C}$ and the enumeration degrees $\mathfrak{E}$, these satisfy $\mathfrak{T} \subsetneq \mathfrak{C} \subsetneq \mathfrak{E} \subsetneq \mathfrak{M}$.

Turing degrees are obtained from the usual Turing reducibility $\leq_T$ defined on points $p, q \in \Baire$ with $p \leq_T q$ iff there is a computable function $F : \subseteq \Baire \to \Baire$ with $F(q) = p$. We thus see $p \leq_T q \Leftrightarrow \{p\} \leq_M \{q\}$, and can indeed understand the Turing degrees to be a subset of the Medvedev degrees. The continuous degrees were introduced by \name{Miller} in \cite{miller2}. Enumeration degrees have received a lot of attention in recursion theory, and were originally introduced by \name{Friedberg} and \name{Rogers} \cite{friedberg}. In both cases, we can provide a simple definition directly as a substructure of the Medvedev degrees later on.

\begin{definition}
For a represented space $\mathbf{X}$, let $$\textrm{Spec}(\mathbf{X}) = \{\delta_{\mathbf{X}}^{-1}(x)/\equiv_M  \mid x \in \mathbf{X}\} \subseteq \mathfrak{M}$$ be the \emph{point degree spectrum} of $\mathbf{X}$.
 \end{definition}

 We observe that if $\mathbf{X} \subseteq \mathbf{Y}$ then $\textrm{Spec}(\mathbf{X})\subseteq \textrm{Spec}(\mathbf{Y})$, and if $\mathbf{X} \cong \mathbf{Y}$, then $\textrm{Spec}(\mathbf{X}) = \textrm{Spec}(\mathbf{Y})$ (the latter making sure that the point degree spectrum is a well-behaved property of represented spaces).

\begin{observation}
Some spectra of known spaces:
\begin{enumerate}
\item $\spec(\Cantor) = \spec(\Baire) = \spec(\mathbb{R}) = \mathfrak{T}$
\item $\spec(\uint^\mathbb{N}) = \spec(\mathcal{C}(\uint,\uint)) =: \mathfrak{C}$

This is essentially the definition of the continuous degrees given by \name{Miller} in \cite{miller2}. As any separable metric space embeds into the Hilbert cube $\uint^\mathbb{N}$, we find in particular that $\spec(\mathbf{X}) \subseteq \mathfrak{C}$ for any computable metric space $\mathbf{X}$.

\item $\spec(\mathcal{O}(\mathbb{N})) =: \mathfrak{E}$

A straight-forward comparison with the usual definition of enumeration degrees shows this. As mentioned above, we shall use this observation as the definition of $\mathfrak{E}$. As any countably based admissible space embeds into $\mathcal{O}(\mathbb{N})$, we find that $\spec(\mathbf{X}) \subseteq \mathfrak{E}$ for all those spaces.
\end{enumerate}
\end{observation}

\begin{theorem}
\label{theo:spectrum}
The following are equivalent for a represented space $\mathbf{X}$:
\begin{enumerate}
\item $\textrm{Spec}(\mathbf{X}) \subseteq \textrm{Spec}(\mathbf{Y})$
\item $\mathbf{X} = \bigcup_{n \in \mathbb{N}} \mathbf{X}_n$ where there are $\mathbf{Y}_n \subseteq \mathbf{Y}$ with $\mathbf{X}_n \cong \mathbf{Y}_n$
\end{enumerate}
\end{theorem}

By combining this with a theorem by \name{Hurewicz} and \name{Wallman} \cite[p50-51]{hurewicz}, we obtain the following, which answer a long-standing question by \name{Pour-El} and \name{Richards} \cite{pourel}:
\begin{corollary}
The following are equivalent for a Polish space $\mathbf{X}$:
\begin{enumerate}
\item $\exists p \ . \ \{p\} \times \textrm{Spec}(\mathbf{X}) \subseteq \mathfrak{T}$
\item $\dim(\mathbf{X}) \neq \infty$
\item $\mathbf{X} = \bigcup_{n \in \mathbb{N}} \mathbf{X}_n$ where $\exists p$ s.t. there are $A_n \subseteq \Cantor$ with $A_n \cong \{p\} \times \mathbf{X}_n$
\item $\mathbf{X} = \bigcup_{n \in \mathbb{N}} \mathbf{X}_n$ where $\exists p$ s.t. there are $A_n \subseteq \Cantor$ with $A_n \cong \{p\} \times \mathbf{X}_n$ and any $A_n$ is $\Pi^0_2$
\end{enumerate}
\end{corollary}

\begin{corollary}[\name{Miller} \cite{miller2}]
$\textrm{Spec}(\mathcal{C}(\uint, \uint)) \nsubseteq \mathfrak{T}$
\end{corollary}

It follows from well-known results that for any uncountable Polish spaces $\mathbf{X}$ we find $\mathfrak{T} \subseteq \textrm{Spec}(\mathbf{X}) \subseteq \mathfrak{C}$.

\begin{problem}[cf.~\cite{schlicht}]\label{prob:third}
Does there exist an uncountable Polish space $\mathbf{X}$ such that
\[\mathfrak{T} \subsetneq \textrm{Spec}(\mathbf{X}) \subsetneq \mathfrak{C} \ \ ?\]
\end{problem}

This question was answered in the affirmative by \name{Kihara} and the author in \cite{pauly-kihara-arxiv} drawing on a variety of results from dimension theory \cite{lelek}, \cite{pol2}, \cite{radul}, \cite{schori}. Recursion theoretic methods produce a much stronger result -- there are even uncountably many distinct point degree spectra of Polish spaces between $\mathfrak{T}$ and $\mathfrak{C}$, including uncomparable ones. Beyond that, the structure of the point degree spectra under inclusion of Polish spaces, quasi-Polish spaces and admissible spaces seems to be mostly unknown, and a promising target for further investigations drawing on both recursion theory and dimension theory.

\section{Additions to the theory of represented spaces}
Various diverse research endeavors recently highlighted the need for additional features in the theory of represented spaces, in particular there seem to be further fundamental properties of represented spaces that are likely to appear in requirements of theorems.

A rather straight-forward property is a generalization of completeness including in the requirements for Polish spaces. As shown by \name{de Brecht} in \cite{debrecht6}, the following also generalizes Smyth-complete quasi-metrizability:

\begin{definition}
We call a represented space $\mathbf{X}$ \emph{complete}, if it admits a total representation $\delta_\mathbf{X} : \Baire \to \mathbf{X}$.
\end{definition}

\name{Selivanov}'s \cite{selivanov5} can be seen as an investigation of some basic aspects of complete represented spaces. A crucial consequence of completeness for descriptive set theory seems to be related to the countable ordinals. Given an appropriate represented space $\textrm{COrd}$ of the countable ordinals, we can prove (similar to \cite[Lemma 26 (5)]{pauly-gregoriades-arxiv}):

\begin{theorem}
Let $\mathbf{X}$ be a complete represented space. If $f : \mathbf{X} \to \textrm{COrd}$ is continuous, then it is bounded, i.e.~$\exists \alpha \in \textrm{COrd} \ \forall x \in \mathbf{X} \ f(x) \leq \alpha$.
\end{theorem}

What exactly is a suitable representation for the countable ordinals, and whether there is a representation that would make it possible to compute a bound from the function in the preceding theorem, is an open question at this stage (and should be one of the next questions to addressed within the proposed programme). Note that while the notion of a \emph{computable ordinal} is rather robust, the various representation implicit in the usual characterizations are not equivalent.

In some cases, a far stronger notion is needed to make a theorem work:
\begin{definition}
We call a represented space $\mathbf{X}$ \emph{completely compact}, if it admits a total representation $\delta_\mathbf{X} : \Cantor \to \mathbf{X}$.
\end{definition}

Every completely compact space is both compact and complete -- the converse seems like a reasonable conjecture, which has however not been resolved so far.

Two further properties concern the complexity of the sets of names of points, and as such behave like converse notions to admissibility to some extent. We call a representation $\delta_\mathbf{X}$ \emph{effectively fiber-compact}, if $\delta_\mathbf{X}^{-1} : \mathbf{X} \to \mathcal{K}(\Cantor)$ is computable, and \emph{effectively fiber-overt}, if $\delta_\mathbf{X}^{-1} : \mathbf{X} \to \mathcal{V}(\Cantor)$ is computable. Effectively fiber-compact representations are required for the approach to complexity theory suggested by \name{Weihrauch} in \cite{weihrauchf}, where is also shown that any computable metric space has an effectively fiber-compact representation. It is also the case that any computable metric space has a fiber-overt representation -- however, in general both properties cannot be attained by the same representation! Effectively fiber-overt representations are required for some constructions used in \cite{paulybrattka4} by \name{Brattka} and the author to make constructions of the represented space of continuous \emph{multi}-valued functions between represented spaces well-behaved.

While it is still open how widespread effectively fiber-overt representations are available, for effectively fiber-compact representations this was classified precisely by \name{Kihara} and the author in \cite{pauly-kihara-arxiv} using a metrization theorem from \cite{grubba3,weihrauchm}:

\begin{theorem}
A represented space $\mathbf{X}$ admits a  computably admissible effectively fiber-compact representation iff $\mathbf{X}$ embeds computably into a computable metric space.
\end{theorem}

\section{The Jayne-Rogers theorem and the representability conjecture}
Call a function $f : \mathbf{X} \to \mathbf{Y}$ $\Pi^0_1$-piecewise continuous, if there is a cover $(A_i)_{i \in \mathbb{N}}$ of $\mathbf{X}$ of closed sets (i.e.~$\Pi_1^0$-sets), such that any $f_{|A_i}$ is continuous.
\begin{theorem}[Jayne \& Rogers \cite{jaynerogers}]
Let $\mathbf{X}$ be Polish and $\mathbf{Y}$ be separable. Then a function $f : \mathbf{X} \to \mathbf{Y}$ is piecewise continuous iff it is $\Delta_2^0$-measurable.
\end{theorem}

The Jayne-Rogers theorem and its potential generalizations have received plenty of attention, e.g.~\cite{ros,ros2,semmes}. Phrased in the terms of synthetic descriptive set theory it takes the form of stating that certain spaces are $d$-admissible for a certain endofunctor $d$, and as such is identified as a parallel result to the Banach-Lebesgue-Hausdorff theorem.

The endofunctor we need to consider is connected to computation with finitely many mindchanges or non-deterministic computation with discrete advice as studied by \name{de Brecht}, \name{Brattka}, \name{Ziegler} and the author in \cite{paulybrattka,paulybrattka2,debrecht,debrecht5,ziegler3,ziegler5,ziegler7}.

Define $\Delta : \subseteq \Baire \to \Baire$ via $\Delta(p)(n) = p(n + 1 + \max \{i \mid p(i) = 0\}) - 1$. Let the finite mindchange endofunctor be defined via $(X, \delta_X)^\nabla = (X, \delta_X \circ \Delta)$ and $(f : \mathbf{X} \to \mathbf{Y})^\nabla = f : \mathbf{X}^\nabla \to \mathbf{Y}^\nabla$. In $^\nabla$ we have a computable endofunctor that preserves binary products, but not countable ones. On Polish spaces, the $^\nabla$-continuous functions are precisely the $\Pi^0_1$-piecewise continuous ones, and the $^\nabla$-open sets are the $\Delta^0_2$-sets. The property of being $^\nabla$-Hausdorff is a uniform version of the $T_D$-separation axiom. In \cite{paulydebrecht}, \name{de Brecht} and the author have shown the:

\begin{theorem}[Computable Jayne-Rogers-theorem \cite{paulydebrecht}]
Let $\mathbf{Y}$ be computably admissible, completely compact and $^\nabla$-Hausdorff. Then $\mathbf{Y}$ is $^\nabla$-Hausdorff.
\end{theorem}

There is a certain catch here, which is that the computable Jayne-Rogers theorem does not imply its classical counterpart. The reason for this that any measurability notion in synthetic descriptive set theory is internal to the setting: The preimage-map will be continuous. In some situations this is unproblematic: Between countably-based spaces $\mathbf{X}$,$\mathbf{Y}$, a preimage function $f^{-1} : \mathcal{O}(\mathbf{Y}) \to \mathcal{O}'(\mathbf{X})$ is automatically continuous whenever it is well-defined. Now the classical Jayne-Rogers-theorem does indeed imply that this remains true for $f^{-1} : \mathcal{O}(\mathbf{Y}) \to \mathcal{O}^\nabla(\mathbf{X})$ whenever these spaces satisfy it -- but a direct proof is unknown! This situation becomes even more aggravating when taking into consideration that \name{Kihara} in \cite{kihara4} was able to prove a computable version of a generalization of the Jayne-Rogers theorem which has long been an unattained goal in classical descriptive set theory. If only we knew that our approach takes into consideration all functions from classical descriptive set theory, this problem would be resolved.

To make these considerations slightly more precise we introduce some notation. Given two computable endofunctors $c$, $d$ and represented spaces $\mathbf{X}$, $\mathbf{Y}$, let $\mathcal{C}_{-1}(\mathcal{O}^c(\mathbf{Y}), \mathcal{O}^d(\mathbf{X})) := \{g \in \mathcal{C}(\mathcal{O}^c(\mathbf{Y}), \mathcal{O}^d(\mathbf{X})) \mid \exists f : X \to Y \ g = f^{-1}\}$ be the space of continuously $c, d$-measurable functions. Given some computable endofunctor $d$, we can obtain a computable endofunctor $\chi^d$ such that the $\chi^d$-continuous maps are those admitting a countable partition into $\mathcal{A}^d$-sets such that any restriction is continuous.

\begin{theorem}[\name{Kihara} \cite{kihara4}]
Let $\mathbf{X}$, $\mathbf{Y}$ satisfy $\operatorname{Spec}(\mathbf{X}), \operatorname{Spec}(\mathbf{Y}) \subseteq \mathfrak{T}$. Let $\alpha$, $\beta$ be countable ordinals with $1 < \alpha \leq \beta \leq 2\alpha$, let $\gamma$ be the smallest ordinal such that $\gamma + 1 + \alpha > \beta$. Then there is a computable bijection between $\mathcal{C}_{-1}(\mathcal{O}^{(\alpha)}(\mathbf{Y}), \mathcal{O}^{(\beta)}(\mathbf{X}))$ and $\mathcal{C}(\mathbf{X}, \chi^{(\beta)}\mathbf{Y}^{(\gamma)})$.
\end{theorem}

\name{Gregoriades} (personal communication) has observed that a result by \name{Louveau} \cite{louveau} can be used to show that any $(\Sigma^0_{n+1}, \Sigma^0_{m+1})$-measurable function $f$ between Polish spaces $\mathbf{X}$, $\mathbf{Y}$ does have a preimage function $f^{-1} \in \mathcal{C}(\mathcal{O}^{(n)}(\mathbf{Y}), \mathfrak{b}\left (\mathcal{O}^{(m)}(\mathbf{X})\right ))$, where $\mathfrak{b}$ is the computable endofunctor capturing Borel measurability.

\begin{conjecture}[Weak representability conjecture]
The underlying sets of $\mathcal{C}_{-1}(\mathcal{O}^{(n)}(\mathbf{Y}), \mathfrak{b}\left (\mathcal{O}^{(m)}(\mathbf{X})\right ))$ and $\mathcal{C}_{-1}(\mathcal{O}^{(n)}(\mathbf{Y}), \left (\mathcal{O}^{(m)}(\mathbf{X})\right ))$ contain the same elements.
\end{conjecture}

\begin{conjecture}[Strong representability conjecture]
The spaces $\mathcal{C}_{-1}(\mathcal{O}^{(n)}(\mathbf{Y}), \mathfrak{b}\left (\mathcal{O}^{(m)}(\mathbf{X})\right ))$ and $\mathfrak{b} \mathcal{C}_{-1}(\mathcal{O}^{(n)}(\mathbf{Y}), \left (\mathcal{O}^{(m)}(\mathbf{X})\right ))$ are computably isomorphic.
\end{conjecture}

Using \name{Kihara}'s result, it follows that the decomposability conjecture (with the corresponding restrictions on the parameters) is equivalent to the weak representability conjecture, which in turn would be implied by the strong representability conjecture. If the latter is true, it does not seem entirely unreasonable to hope for an elegant proof thereof.
\section{Further connections and questions}
\subsection{Falsifiability in the philosophy of science}
\name{Popper} famously used the notion of \emph{falsifiability} in his attempt to formulate a both normative and descriptive theory of science \cite{popper2}. In a synthetic reading, a falsifiable property is nothing but a closed set. A prominent concern in the reception of \name{Popper}'s theory (e.g.~\cite{feyerabend,andersson}) then essentially boils down to the observation that not all sets of interest for science are necessarily either closed or open, with concrete examples usually being $\Sigma^0_2$ or $\Pi^0_2$ sets.

A descriptive set theory freed from its constraint to Polish spaces may be very amenable for a deepened interaction with the philosophy of science. Moving the focus away from the logical form of a statement to a more process-orientated view as how to ascertain its truth (which is what underlies the approach via endofunctors) is probably beneficial, too.

\subsection{Infinite games and Borel determinacy}
Consider a full infinite binary tree, where each vertex is controlled by one of two agents. By choosing successors for the vertices they control, the two agents jointly pick an infinite path $p$ through the tree. The game is further described by some set $W \subseteq \Cantor$: If $p \in W$, the first agent wins, if $p \notin W$ the second agents wins. A celebrated result \cite{martin} by \name{Martin} states that whenever $W$ is a Borel set, then either the first player has a strategy that forces the play to fall into $W$ no matter what the second player does, or the second player has a strategy enforcing $p \notin W$ regardless of the first players actions.

Our more general framework now allows us to compare the complexity of the set $W$ describing the game with the complexity of the mapping that takes the description of the game and returns a winning strategy for the appropriate player. Preliminary work by \name{Le Roux} and the author \cite{paulyleroux3-arxiv} shows that if the former climbs up the difference hierarchy, the latter climbs up the Borel hierarchy. The current proofs are rather technical (partly reusing constructions from \cite{nemoto} by \name{Nemoto}, \name{MedSalem} and \name{Tanaka}) -- it would be very interesting to find more abstract proofs, maybe even synthetic ones!

\subsection{Connections to Weihrauch reducibility}
The Weihrauch degrees from the framework for the research programme to classify the computational content of mathematical theorems formulated by \name{Brattka} and \name{Gherardi} \cite{brattka3} (also \name{Gherardi} \& \name{Marcone} \cite{gherardi}, \me \cite{paulyincomputabilitynashequilibria}). The core idea is that $S$ is Weihrauch reducible to $T$, if $S$ can be solved using a single invocation of $T$ and otherwise computable means. The examples of computable endofunctors studied so far are all derived (via the concept of a jump operator, \name{de Brecht} \cite{debrecht5}) from well-studied Weihrauch degrees. Results in \cite{paulybrattka4} suggest a rather general procedure to move from a Weihrauch degree to a computable endofunctor.

The numerous theorems that have been classified would then always yield natural examples of maps complete for $d$-continuity for a certain $d$. Some example are the separable Hahn-Banach theorem (\name{Gherardi} \& \name{Marcone} \cite{gherardi}), the Intermediate Value Theorem (\name{Brattka} \& \name{Gherardi} \cite{brattka3}), Nash's theorem for bimatrix games (\me \cite{paulyincomputabilitynashequilibria}), Brouwer's Fixed Point theorem (\name{Brattka}, \name{Le Roux} \& \me \cite{paulybrattka3cie}), the Bolzano-Weierstrass theorem (\name{Brattka}, \name{Gherardi} \& \name{Marcone} \cite{gherardi4}), the Radon-Nikodym derivative (\name{Hoyrup}, \name{Rojas} \& \name{Weihrauch} \cite{hoyrup2b}), the Lebesgue Density Lemma (\name{B.}, \name{Gherardi} \& \name{H\"olzl} \cite{hoelzl}), the Goerde-Browder-Kirk fixed point theorem (\name{Neumann} \cite{eike-neumann}) and the Frostmann lemma (\name{Fouch\'e} and \me \cite{pauly-fouche}).

As already noticed by \name{de Brecht} in \cite{debrecht5}, it seems reasonable to expect that also the structure theory of Weihrauch degrees can be useful starting point to understand the structure theory of computable endofunctors. This should be rather straight-forward for the algebraic structure theory investigated by \me \cite{paulymaster,paulyreducibilitylattice}, \name{Brattka} \&. \name{Gherardi} \cite{brattka2}, \name{Higuchi} and \me \cite{paulykojiro} and \name{Brattka} and \me \cite{paulybrattka4}. However, also somewhat more idiosyncratic structure results such as the fractal absorption theorems from \cite{paulyleroux-arxiv} by \name{Le Roux} and \me maybe useful -- in particular seeing that the proof of the computable Jayne-Rogers theorem by \me and \name{de Brecht} \cite{paulydebrecht} is using a very similar theorem from \cite{paulybrattka} by \name{Brattka}, \name{de Brecht} and the author.
\bibliographystyle{eptcs}
\bibliography{../../spieltheorie}

\section*{Acknowledgements}
The work presented here benefited from the Royal Society International Exchange Grant IE111233 and the Marie Curie International Research Staff Exchange Scheme \emph{Computable
Analysis}, PIRSES-GA-2011- 294962.

Ideas rarely appear out of nothing, and I am grateful to Vasco Brattka, Martin Escard\'o, Vassilios Gregoriades, T\'amas Kisp\'eter, St\'ephane Le Roux, Zachiri McKenzie, Luca Motto-Ros, Eike Neumann, Matthias Schr\"oder and Victor Selivanov for discussions conductive to the inception of the research programme outlined here. Special thanks go to Matthew de Brecht and Takayuki Kihara, my coauthors on \cite{paulydebrecht,paulydebrecht2} and \cite{pauly-kihara-arxiv} respectively, without whom this research would have been impossible.
\end{document}